\documentclass[12pt]{article}
\begin{document}
\begin{center}
{\large\bf New measures \\
of the quality and of the reliability of fits \\ applied
to forward hadronic data at $t=0$
\footnote{\rm presented by J.R. Cudell
at the 6$^{th}$ workshop on non-perturbative QCD, American University of Paris,
5-9 June 2001.}}
\vskip 1cm

{J.R. Cudell} 

{\it \small Inst. de Physique, B\^at. B5,
Univ. de Li\`ege, Sart Tilman, B4000 Li\`ege, Belgium  
}
~\\~\\

{ K. Kang}

{\it \small Physics Department, Brown University,
Providence, RI, U.S.A.  }
~\\~\\

{V.V. Ezhela, Yu.V. Kuyanov,  S.B. Lugovsky,  N.P. Tkachenko}

{\it \small COMPAS group, IHEP, Protvino,
Russia }  
~\\~\\

{ P. Gauron, B. Nicolescu}

{\it \small LPTPE, Universit\'e Pierre et Marie
Curie, Tour 12 E3, 4 Place Jussieu, 75252 Paris Cedex 05, France}
~\\~\\

{(COMPETE collaboration)}
\end{center}
\vskip 2cm
\begin{center}
{\bf Abstract}
\end{center}
{We develop five new statistical measures of the quality
of fits, which we combine with the usual confidence level to determine
the models which fit best all available data for total cross sections
and for the real part of the forward hadronic amplitude.}
\newpage
Phenomenological studies compare models with data,
in order to determine whether these agree together. The typical criterion
is that there is good agreement if the $\chi^2$ per degree of freedom ($\chi^2/dof$)
is of the order of 1, or if the confidence level is bigger than a typical
value. To take an explicit example, one can imagine fitting
all available elastic hadron-hadron amplitudes (their imaginary part being provided
by the total cross sections, and their real part by the $\rho$ parameters)
to analytic parametrisations\cite{Cudell:2001}. Although we shall only 
treat this case explicitly,
the remarks and tools given here do apply to most situations.


There are several problems with this conventional approach: first of all, models usually
do not apply everywhere, hence the comparison should hold only for part of the
data. In the case of hadronic amplitudes, the models are smooth analytic 
functions, which are expected to work at high enough energy. Some of 
these models are also expected to have large (unitarising) corrections 
at large energies.  Hence in this simple case, there is a range of energy 
over which the comparison should be meaningful. However, as is often the case, 
the exact range over which the models hold is not predicted by
the theory, and should emerge from the fits themselves.

Secondly, once one fits many points, a large discrepancy between the theory
and a few of the data points can be overshadowed by a good overall agreement.
This has the drawback that it is precisely these points which may point out to
new physics, but at the same time this may reveal problems with the data.
Hence some uniformity in the description of the data is needed, and studies
such as the present one, applied to soft hadronic amplitudes, lead to a reassessment 
of the data used in the fit.

Thirdly, the fits to models lead to some values for the physical parameters of the model.
Here, one must take into account the interplay between these values and 
the data sub-sample for which the model applies. This means that if one changes that sample
slightly, the parameters extracted for the models should be stable. A typical counter-example\cite{Cudell:2000}
is found in the fits to a simple-pole pomeron, which give (wrongly) an increasing pomeron intercept once the
minimum energy of the data is below 9 GeV. 

Because we want to be able to consider a large data sample and many possible models,
we are aiming at the development of an automatic decision-making procedure, and hence 
we want to $measure$ the above criteria. Although we are not entirely finished with
this program, we can present what seems a reasonable set of measures
which reflect
the above aspects of the fits.  All these measures, or {\it indicators},  are
constructed so that the higher their value the better is the quality
of the data description.

The first indicator concerns the sample of data that can be fitted.
In the case of hadronic amplitudes, we shall consider the range of energies where 
a given model applies, defined as the region in energy
where the fit has a confidence level ($CL$) bigger than 50\%.
Its size will be one of the measures of the quality of the fit: we define the
{\bf applicability} $A$ of model $M$ as: 
\begin{equation}
\label{A}
A_{j}^{M}=w_{j} \log \left({E^{M,high}_{j}
\over E^{M,low}_{j}}\right),\ \ A^{M}={ \sum _{j}A^{M}_{j}\over N_{sets}}
\end{equation}
where \( E^{M,high}_{j} \) ( resp. \( E^{M,low}_{j} \) ) are respectively
the highest and lowest values of the energy
in the area of applicability of model \( M \) in the data subset
\( j \) 
and \( w_{j} \) is the weight determined from the best
fit in the same interval. 
After we have defined where the model may work, we can check how well
it fits, although by definition all models will provide a satisfactory fit. 
We may consider the usual confidence level, 
$ C^{M}_{1}=CL(\%)$, where the $CL$ refers to the whole area of applicability of the model $M$,
or a reduced one 
$ C^{M}_{2}$ limited to the intersection of the areas of applicability
of all models qualified for the comparison.

The next measure of quality has to do with the number of parameters of the model,
given the number of data points in the range of applicability. Hence we
define the {\bf rigidity} $R_1$ as:
\begin{equation}
\label{R1}
R^{M}_{1}={N_{dp}^{M}(A)\over {1+N^{M}_{par}}}
\end{equation}

All the information on a given fit is contained in
the error matrix, and we shall use it to define the new measures. Hence the
first condition to check is whether the error matrix itself is reliable, $i.e.$
whether the correlations between parameters are minimal. Hence we define the 
{\bf reliability} $R_2$ as:
\begin{equation}
\label{R2}
R^{M}_{2}={2\over N_{par}(N_{par}-1)}\cdot \sum _{i>j=1}^{N}\Theta (90.0-C^{M}_{ij})
\end{equation}
 where \( C^{M}_{ij} \) is the correlation matrix element in \( \% \)
calculated in the fit at the low edge of the applicability area.

We are now in a position to define the stability of the model with respect to the 
data range considered. In the case of hadronic amplitudes, three possible changes
can be considered: we can consider the variation that comes from modifying the energy 
threshold of the fit ({\bf energy stability} $S_1$), or the fluctuation of the $\chi^2$
from bin to bin for some data binning
motivated by physics (in the case of hadronic amplitudes, we bin according
to the process/observable) 
({\bf uniformity} $U$), or  the reproducibility
of the parameters values when fitting, with the same number of adjustable
parameters, a reduced data sample and
a reduced number of observables,
in the case of hadronic amplitudes when excluding the real part 
({\bf r-stability} $S_2$).
 The latter is introduced in this case because the data for $\rho$ parameter data may 
be less reliable than those for the cross section. 
Hence we obtain the three measures:
\begin{eqnarray}
{1\over S^{M}_{1}}&=&{1\over N_{steps}N_{par}^{M}}\sum _{steps,ij}(P^{t}-P^{step})_{i}(W^{t}+W^{step})^{-1}_{ij}(P^{t}-P^{step})_{j}\nonumber\\
{1\over U^{M}}&=&{1\over N_{sets}}\sum _{j}{1\over 4}\left[ \frac{\chi^{2}(t)}{N^{t}_{nop}}-\frac{\chi^{2}(j)}{N^{j}_{nop}}\right] ^{2}\\
{1\over S^{M}_{2}}&=& {1\over 2N_{par}^{M}}\sum _{ij}(P^{t}-P^{t(no\: \rho
)})_{i}(W^{t}+W^{t(no\: \rho )})^{-1}_{ij}(P^{t}-P^{t(no\: \rho )})_{j}
\label{S1}
,\nonumber\end{eqnarray}
 where: \( P^{t} \) is the vector of parameters values obtained from the
model fit to the whole area of applicability;
\( P^{step} \) is the vector of parameters values obtained from the model
fit to the reduced data set on the \( step \), in our case \( step \)
means a shift in the low edge of the fit interval to the right by 1
GeV;
\( W^{t} \) and \( W^{step} \) are the error matrix estimates obtained
from the fits to the total and to the reduced 
data samples from the domain of applicability, 
\( t \) denotes the total area of applicability,
and $t(no\ \rho)$ the data sample with $\rho$ data excluded.

Having defined these measures of the quality of fits, we want to use them
to see whether we can decide which is the safest model to use to reproduce 
a given set of data. As already emphasized\cite{Cudell:2000} in the case of 
hadronic amplitudes, 
standard methods do not allow one to decide which models are 
to be preferred, and
several classes of parametrisations are possible. Using these new measures,
we can try to decide which models are best. All models considered are the
sum of several terms: the low-energy sector is described by an amplitude with
the following imaginary part, with $s_1=1$ GeV$^2$:
\begin{equation}
Im(A^{ab})=Y^{ab}_1 \left({s/s_1 }\right)^{\alpha_1}
\mp Y^{ab}_2 \left({s/s_1 }\right)^{\alpha_2}
\end{equation}
The first term has charge-conjugation $C=+1$ whereas the second has \break
$C=-1$ (with
the $-$ sign for a positively charged beam). 
These two terms are symbolized by the notation $RR$ in the following.
The high-energy behaviour is dominated by a pomeron term, for which we 
consider the following terms, or their combination, in the imaginary part 
of the amplitude:
\begin{eqnarray}
Im(A^{ab})&=&X^{ab} \left({s/s_1 }\right)^{\alpha_{\wp}}\\
Im(A^{ab})&=&Z^{ab} s\\
Im(A^{ab})&=&B^{ab} s\ln\left({s/s_1}\right)\\
Im(A^{ab})&=&B^{ab} s\ln^2\left({s/s_1}\right)
\end{eqnarray}
which we denote respectively by $E$, $P$, $L$ and $L2$. 
If we take for the pomeron
a simple pole model we obtain in this notation $RRE$, whereas a double pole
gives $RRPL$ and a triple pole $RRPLL2$, which we can write as $RRPL2$ with
a scale $s_0$ instead of $s_1$ in the $\log^2$.

Furthermore, we have considered several possibilities to constrain
the parameters. The following notations are attached as either superscript
or subscript to the model variants in each case:\\
{\bf{d}} means degenerate leading reggeon trajectories \( \alpha
_{1}=\alpha _{2} \);\\
{\bf{u}} means universal (independent of projectile hadron);\\
{\bf{nf}} means that we have not imposed factorization for the residues
of the pomeron term(s) in the case of the \( \gamma \gamma  \)
and \( \gamma p \) cross sections;\\ 
{\bf{qc}} means that a quark counting rule is imposed
on the residues of the amplitude for $\Sigma p$ scattering, constrained by
the residues in $pp$ and $Kp$;\\
{\bf c} implies the use of the Johnson-Treiman-Freund 
relation for the cross section differences: \( \Delta \sigma (N)=5\Delta \sigma
({\pi }),\Delta \sigma (K)=2\Delta \sigma ({\pi }) \).\\
Finally, the real parts of the amplitudes can be obtained through 
$s\rightarrow u$ crossing.

The first results are quite generic and based on a study of the $\chi^2$ alone:
\begin{description}
\item{(1)} All analytic descriptions of the data based on the above terms 
break down at $\sqrt{s}\leq 4$ GeV;
\item{(2)} Most models require a non-degeneracy of lower trajectories. 
Degeneracy can be accommodated only by $RRPL2$;
\item{(3)} Simple pole pomerons fail to reproduce the real part of the cross
section, and all models have problems in reproducing some of the $\rho$ data;
\item{(4)} Cosmic ray data are well reproduced by the best parametrisation, 
with no need of re-analysis of the published data;
\item{(5)} Although quark counting rules can be approximately implemented, it
is also possible to have a universal rising term for the pomeron\cite{Gauron:2000}.
\end{description}
Furthermore, we can use all the information contained in our indicators to
define the best models. Several schemes are possible:\\
i) the ACCURRSS scheme: we take all indicators, including the $C_1$ and $C_2$,
and for each indicator we order the $N$ models considered from rank 1 to $N$ 
according to the value of the indicator. We then sum the 8 numbers obtained, 
and the best model is the one with highest  rank overall;\\
ii) As the indicators are statistical measures, we can do the same as above
but consider that a model is better than another (and give it one point) 
only if its indicator is bigger than that of the other model by {\it e.g.} 20\%. 
This leads to the ACCURRSS$_{20}$
scheme;\\
iii) Finally, one may argue that all the $CL$ are acceptable, and that the
number of parameters is not relevant as we have chosen functional forms,
and hence in principle an infinite number of parameters. Using a statistical
ranking similar to ii) leads then to the AURSS$_{20}$ scheme.

\begin{table}[t]
\caption{Best models for total cross sections
}
\begin{center}
\footnotesize
\begin{tabular}{|c|c|c|c||}
\hline
 & {ACCURRS} &ACCURRS$_{20}$&AURS$_{20}$\\
\hline
1&  $RRL2_{qc}$          &$R_{qc}R_cL2_{qc}$    &$R_{qc}R_cL2_{qc}$  \\
2&  $(RR)_dPL2_u$          &$(RR_c)_dPL2_u$       &$(RR_c)_dP_{qc}L2_u$\\
3&  $(RR_c)_dPL2_u$      &$(RR)_dPL2_u$         &$R_{qc}R_cL_{qc}$    \\
4&  $ R_{qc}R_cL2_{qc}$  &$RRL2_{qc}$           &$(RR_c)_dPL2_u$ \\
\hline
\end{tabular}
\end{center}
\end{table}
Using these ranking schemes, we obtain the best models (out of the order of
30 variations on the terms used in the models) given in Table 1 for 
fits to total cross sections only, and in Table 2 for fits to all data for 
hadronic amplitudes. As can be seen from these table, simple-pole pomerons are 
never preferred, and models containing a $\log^2 s$ rise in the cross section 
always provide the best fits to the data. To reach a more restrictive conclusion,
one needs to use the $\rho$ data, in which case the preferred model is
consistently $RRPL2_u$. The problem however is that the $\rho$ data are 
poorly reproduced by
all models considered in this study, hence one cannot be sure that this 
preference will survive future iterations of the cross 
assessments with new models and new data added,
such as off-forward cross sections, or DIS structure functions, or data from 
future experiments. As is always
the case in these studies, it would be of utmost interest to have higher-energy
data for other beams than protons and antiprotons. This would of course 
enable one to determine directly 
whether the pomeron counts quarks, or has a universal
component.
\begin{table}[t]
\caption{Best models for hadronic amplitudes 
}
\begin{center}
\footnotesize
\begin{tabular}{|c|c|c|c|}
\hline
 & {ACCURRSS} &ACCURRSS$_{20}$&AURSS$_{20}$\\
\hline
1& $RRPL2_u$        &$RRPL2_u$                     &$RRPL2_u$  \\
2& $RRL_{nf}$        &$RR_cL2_{qc}=RRL2_{qc}$             &$RRL2 $       \\
3& $(RR_c)_dPL2_u$  &                            &$R_{qc}R_cL2_{qc} $     \\
4& $(RR)_dPL2_u$   &$(RR)_dPL2_u=R_{qc}R_cL2_{qc}=RRL2$  &$(RR)_dPL2_u$  \\
\hline
\end{tabular}
\end{center}
\end{table}
\section*{Acknowledgments}
COMPAS was supported in part by the Russian Foundation for
Basic Research grants RFBR-98-07-90381 and RFBR-01-07-90392. K.K.
is in part supported by the U.S. D.o.E. Contract DE-FG-02-91ER40688-Task
A.  We thank the president C.W. Kim of the Korea Institute for Advanced 
Study, Yonsei University and Professor J.-E. Augustin of
LPNHE-University Paris 6 for their hospitality during various stages of this 
work, which made it possible.

\end{document}